\documentclass[aps,prl,twocolumn,superscriptaddress,showpacs]{revtex4-1}

\usepackage{float}
\usepackage{graphicx}
\usepackage{amsmath,amssymb}
\usepackage{color}
\usepackage{framed}
\usepackage{mathtools}
\renewcommand{\vec}[1]{\mathbf{#1}}

\def\v{p}
\def\Np{N_P}
\def\xd{\dot{x}}
\def\qm{\frac{q}{m}}
\def\xdd{\ddot{x}}
\def\xddd{\dddot{x}}

\def\np{401}
\def\SI{19.9}
\def\initialerror{0.5}
\def\cp{465}
\def\cs{5.19}
\def\cS{-1.0}
\def\qap{574}
\def\qas{8.24} %8.2422
\def\qaS{-0.58} %-0.57998
\def\qbp{904}
\def\qbs{14.5}
\def\qbS{-0.16}
\def\qcp{1232}
\def\qcs{17.8}
\def\qcS{-0.041}

\begin{document}

\title{Semi-classical beam cooling in an intense laser pulse}

\author{S.~R. Yoffe}
\affiliation{Department of Physics, SUPA, University of Strathclyde, Glasgow
G4 0NG, United Kingdom}%

\author{Y. Kravets}% etc
\affiliation{Centre de Physique Th\'{e}orique,
\'{E}cole Polytechnique, 91120, Palaiseau, France}

\author{A. Noble}
\email{adam.noble@strath.ac.uk}
\affiliation{Department of Physics, SUPA, University of Strathclyde, Glasgow G4 0NG, United Kingdom}%

\author{D.~A. Jaroszynski}%
\email{d.a.jaroszynski@strath.ac.uk}
\affiliation{Department of Physics, SUPA, University of Strathclyde, Glasgow G4 0NG, United Kingdom}%

\date{\today}

\begin{abstract}
We present a novel technique for studying the evolution of a particle
distribution using single particle dynamics such that the distribution can
be accurately reconstructed using fewer particles than existing
approaches. To demonstrate this, the Landau--Lifshiftz description
of radiation reaction is adapted into a semi-classical model, for which the
Vlasov equation is intractable. Collision between an energetic electron
bunch and high-intensity laser pulses are then compared using the two
theories. Reduction in beam cooling is observed for the semi-classical case.
\end{abstract}

% insert suggested PACS numbers
%\pacs{}
%\keywords{}

\maketitle

The emergence over the next few years of a new generation of ultra-high
power laser facilities, spearheaded by the Extreme Light
Infrastructure \cite{url_ELI} (ELI), represents a major advance in the possibilities afforded by laser
technology. As well as important practical applications, these facilities
will for the first time
allow us to probe qualitatively new physical
regimes. One of the first effects to be explored will be radiation
reaction.

Radiation reaction---the recoil force on an electron due to its emission of
radiation---remains a contentious area of physics after more than a century
of investigation. The standard equation describing radiation reaction (the
so-called LAD equation, after its progenitors Lorentz, Abraham, and Dirac
\cite{Lorentz1916,Abraham1932,Dirac1938})
for a particle \footnote{Typically, radiation reaction effects will be most
  prominent for electrons, for which $q=-e$ and $m=m_e$.} of mass $m$ and
  charge $q$ in an electromagnetic field $F$ reads
\begin{equation}
\label{LAD}
\xdd^a=-\qm F^a{}_b \xd^b+ \tau \Delta^a{}_b \xddd^b.
\end{equation}
Here $\tau \coloneqq q^2/6\pi m$ is the `characteristic time' of the particle
($\simeq6\times 10^{-24}$~s for an electron);
$\Delta^a{}_b \coloneqq \delta^a_b+\xd^a\xd_b$ is the $\xd$-orthogonal projection;
and an overdot denotes differentiation with respect to proper time. Indices
are raised and lowered with the metric tensor $\eta=\text{diag}(-1,1,1,1)$,
and repeated indices are summed from 0 to 3. We work
in Heaviside-Lorentz units with $c=1$.

Despite numerous independent derivations of equation \eqref{LAD}
\cite{Dirac1938,Bhabha1939,Barut1974},
it is subject to numerous difficulties;
see the recent review \cite{Burton2014} for an account of these problems,
and some of the proposed solutions. The most widely used alternative is that
introduced by Landau and Lifshitz \cite{Landau1962}, by treating the self-force as a small
perturbation about the applied force:
\begin{align}
 \label{LL}
 \ddot{x}^a = -\frac{q}{m}F^{ab}\dot{x}_b - \tau \frac{q}{m} \left(
 \dot{F}^{ab}\dot{x}_b - \frac{q}{m}\Delta^a{}_b F^{bc} F_{cd} \dot{x}^d
\right).
\end{align}
It is often claimed that \eqref{LL} is valid provided only that quantum
effects can be ignored, and though a rigorous demonstration remains elusive
there is mounting evidence that this is indeed the case \cite{Kravets2013a}.

Under the conditions expected at ELI, the caveat `provided that quantum
effects can be ignored' is pertinent. Quantum effects are typically
negligible if the electric field observed by the particle is much less than
the Sauter-Schwinger field \cite{Sauter1931,Schwinger1951} typical of QED processes,
\begin{equation}
\chi \coloneqq \frac{e\hbar}{m^2_e}\sqrt{F^{ab}F_{ac}\xd_b\xd^c}\ll 1.
\end{equation}
For 1 GeV electrons in a laser pulse of intensity $10^{22}$~W/cm$^2$
(parameters typical of ELI), $\chi\sim 0.8$ and quantum effects cannot be
ignored. A complete QED treatment of radiation reaction is problematic to
define but, provided $\chi$ remains less
than about unity, a semi-classical modification to \eqref{LL} should be
valid \cite{Erber1966}.

It is generally accepted that (classical or quantum) radiation reaction
effects will be more readily observed in the behaviour of particles than in
  the radiation they emit \cite{Thomas2012,Ilderton2013}. As such, it is important to be able to accurately
determine the distribution of a bunch of particles evolving according to \eqref{LL} or
its semi-classical extension. Usually this would involve solving a Vlasov
equation or following the evolution of very large numbers of particles,
either of which is computationally very intensive.

An important difference between the classical and quantum picture of
radiation emission by a charged particle can be seen in the radiation spectrum.
Classically, a charged particle can radiate at all frequencies. However,
according to the quantum picture, the energy of the emitted photons is
limited by the energy of the particle. This suppresses emission at high
frequencies, and introduces a cutoff to the radiation spectrum
\cite{Erber1966}.
As such, it is expected that the
effects of radiation reaction are actually over estimated by classical
theories in regimes where quantum effects become important
\cite{Ritus1979}.

In order to account for this reduction of radiation reaction effects in the
Landau--Lifshitz equation of motion, following Kirk, Bell and Arka
\cite{Kirk2009} we scale the radiation reaction force
(the term in parentheses)
in \eqref{LL} by
a function $g(\chi)$.
The full expression for $g(\chi)$ involves a non-trivial integral over
Bessel functions of the second kind. Instead, a fit to data obtained from
this function was found by Thomas \textit{et al.} \cite{Thomas2012} to be
\begin{equation}
\label{eq:QM_g}
g\left(\chi\right) = \left(3.7\chi^3 + 31\chi^2 + 12\chi +1\right)^{-4/9}, 
\end{equation}
and it is this model which we adopt here.
It can be clearly seen that, as $\chi \to 0$, we have
$g\left(\chi\right) \to 1$, therefore recovering the classical
equation of motion. This model essentially reduces
to a rescaling of the characteristic time of the electron, $\tau \to 
g(\chi)\tau$. For $\chi \gtrsim 1$, the stochasticity of
quantum emission becomes important, and the semi-classical model is no
longer applicable \cite{Blackburn2014}.

The motion of a single charged particle colliding with a laser pulse in the
radiation reaction regime has been well studied
\cite{DiPiazza2008,Lehmann2011,Harvey2011}. Instead of having to solve a
Vlasov-type equation on the phase space, we propose
a novel numerical method which allows for the dynamics of a particle
\emph{distribution} to be explored using single-particle equations of
motion such that the distribution can be cleanly and efficiently
reconstructed. While
this approach is quite general and could be used for a variety of systems,
we here consider a distribution of particles subject to equation
\eqref{LL} and its semi-classical extension, without particle-particle
interactions \footnote{For a highly
relativistic particle bunch, these interactions can be
neglected on the time scale of the laser interaction.}. The
Vlasov equation for the latter system has no analytical solution, and would
require significant computing resources to solve numerically.

Assuming that the laser pulse can be approximated by a plane wave with
compact longitudinal support, any spatial spread in the particle distribution would only define the moment in time
when each particular particle enters the pulse. For simplicity, we therefore
take all particles to originate from the same point in space. This is reasonable as
we are primarily interested in the momentum distribution. This also allows
us to consider the particle distribution as a function of the phase $\phi =
\omega t - \vec{k}\cdot\vec{x}$.

Since our pulse is modelled by a plane wave and we focus on the longitudinal
properties of the distribution, we consider the
momenta to be strongly peaked in the transverse directions. As such,
the initial distribution can be taken to be a thermal Maxwell--Boltzmann particle distribution
for the (longitudinal) momentum $\v$ (in units of $mc$)
\begin{equation}
 f\left(\phi=0,\v\right) = \frac{\Np}{\sqrt{2\pi\theta}}
 \exp\left[-\frac{(\v-\bar{\v})^2}{2\theta}\right],
\end{equation}
with the thermal spread $\theta = \mathrm{k}T/mc^2$, where
$\mathrm{k}$ is the Boltzman constant and $\Np$ is the number of particles.
We stress that this initial distribution is chosen for its
simplicity; alternative distributions could be used where appropriate.

Typically, one would sample the distribution at
random, which would require a large number of particles to accurately
represent the distribution. Instead, since the particle number is simply
 $\Np = \int_{-\infty}^{\infty} d\v\ f\left(\phi,\v\right)$,
we determine the momentum spacing
$\delta \v$ between the particles from the initial distribution by
truncating the integral,
so that the particle
number increases by unity in the given momentum interval:
\begin{equation}
\label{eq:momentum_interval}
 1 = \int\limits_{\v-\frac{\delta \v}{2}}^{\v+\frac{\delta \v}{2}}
 d\v\ f\left(0,\v\right) \simeq f\left(0,\v\right)\delta \v.
\end{equation}
This leads to a set of $N_p = 2N_c + 1$ initial momenta $V(0) = \big\{\v_i(0)\big\}$ for $i
\in [-N_c,N_c]$, with
the $\v_i$ generated iteratively from $\v_0 = \bar{\v}$ and $\v_{\pm 1} = \bar{\v} \pm 1/f(0,\bar{\v})$ using
\begin{equation}
 \v_i = \v_{i-2\xi} + \frac{2\xi}{f(0,\v_{i - \xi})}\ \quad\text{with}\quad 
 \xi = \text{sgn}(i).
\end{equation}

As the evolution proceeds, this procedure is applied in reverse to
reconstruct the distribution. The set of momenta $V(\phi)$ is
ordered such that $\v_{i+1} \ge \v_{i}$ and used to find $\delta \v_i(\phi) =
\big(\v_{i+1}(\phi) - \v_{i-1}(\phi)\big)/2$. The velocity distribution is
then defined to be
\begin{equation}
 \label{eq:reconstruct_f}
 f(\phi,\v_i) \coloneqq \frac{1}{\delta \v_i(\phi)}.
\end{equation}
Reconstruction of a distribution from a particle sample can be problematic,
but in this formalism it becomes quite natural.

It has recently been shown using an analytic solution to the Vlasov equation
including radiation reaction according to the Landau--Lifshitz theory
\cite{Noble2014}
that collision with a high-intensity laser pulse leads to a significant
contraction of the particle phase space, resulting in a reduction in the
relative momentum spread. As previously observed, this beam cooling
depends only on the total fluence of the pulse, rather than its duration or
peak intensity independently \cite{Neitz2014}.

For the classical Landau--Lifshitz theory, agreement between this Vlasov solution
and numerical results obtained using the proposed method for sampling and
reconstructing the distribution as discussed above is excellent. However,
for the semi-classical extension, the Vlasov equation is no longer tractable. To demonstrate
the use of our proposed method in such a case, we consider the importance of
quantum effects in the interaction of an electron bunch with a laser pulse
modelled by a plane wave.

In order to establish the impact of radiation reaction on the evolution of the particle distribution
function, and the consequence of allowing the radiation reaction force to be
reduced by the quantum model, we introduce
the relative momentum spread and the momentum skewness (calculated from the
mean $\bar{\v}$ and variance $\theta$):
\begin{align}
 \hat{\sigma}(\phi) = \frac{\sqrt{\theta(\phi)}}{\bar{\v}(\phi)} \quad
 \text{and} \quad S(\phi) = \frac{\left\langle \big[\v-\bar{\v}(\phi)\big]^3
 \right\rangle}{\theta^{3/2}(\phi)}.
\end{align}
The former gives a measure of the beam quality, whereas the latter indicates
how symmetric the distribution is
about its mean. (See also the Supplemental Material.)

We introduce the (null) wave vector $k$ such that (with our choice of
metric) the phase of the pulse may be
expressed as
 $\phi = -k \cdot x = \omega t - \vec{k}\cdot\vec{x}$.
The orthogonal
vectors $\epsilon,\lambda$ satisfy
 $\epsilon^2 = \lambda^2 = 1$ and $k \cdot \epsilon = k \cdot \lambda =
 \epsilon \cdot \lambda = 0$,
and together with $k$ and the null vector $\ell$ (defined to satisfy
$\ell\cdot\epsilon = \ell\cdot\lambda = 0$ and $k \cdot \ell = -1$) form a
basis.

For a plane wave with arbitrary polarisation, the electromagnetic field
tensor $F$ takes the form
\begin{equation}
\frac{q}{m}F^a{}_b = a_\epsilon(\phi) \big(\epsilon^a k_b - k^a \epsilon_b
\big) + a_\lambda(\phi) \big(\lambda^a k_b - k^a \lambda_b \big),
\end{equation}
where the functions $a_{\epsilon,\lambda}(\phi)$ are dimensionless
measures of the electric field strength in the $\epsilon$,$\lambda$
direction. We restrict our attention to a linearly polarised
$N$-cycle pulse, modulated
by a $\sin^2$-envelope \cite{Kravets2013a}, with
\begin{equation}
 a(\phi) = \left\{
   \begin{array}{lll}
     a_0 \sin(\phi) \sin^2(\phi/ 2N) & \ & \text{for }0 < \phi
     <2 \pi N \\
     0 & & \text{otherwise}
   \end{array} \right. ,
\end{equation}
where $a_0$ is the dimensionless (peak) intensity parameter (the so-called
``normalised vector potential'').
This pulse shape offers compact support, allowing the particles to begin and
end in vacuum. The total fluence contained by the pulse is proportional to
\begin{equation}
 \mathcal{E} = \int_0^{2\pi N} d\phi\ a^2(\phi) = \frac{3\pi}{8} N a_0^2.
\end{equation}
In this work, $\mathcal{E}$ is kept constant, which fixes $a_0$ for each
value of $N$. It has been shown \cite{Neitz2013,Noble2014} that the classical
Landau--Lifshitz prediction for the final state of a particle distribution
emerging from the pulse is completely determined by the fluence, whereas
quantum effects are expected to depend on the value of $a_0$ itself. We are
then able to explore the impact of the reduced emission in the
quantum model with varying $a_0$ while maintaining the same classical prediction. This allows
us to comment on the \emph{relative} importance of the quantum effects.

\begin{figure*}[tb]
 \begin{center}
  \includegraphics[width=0.94\textwidth, clip=true, trim=0 70 0 0]
  {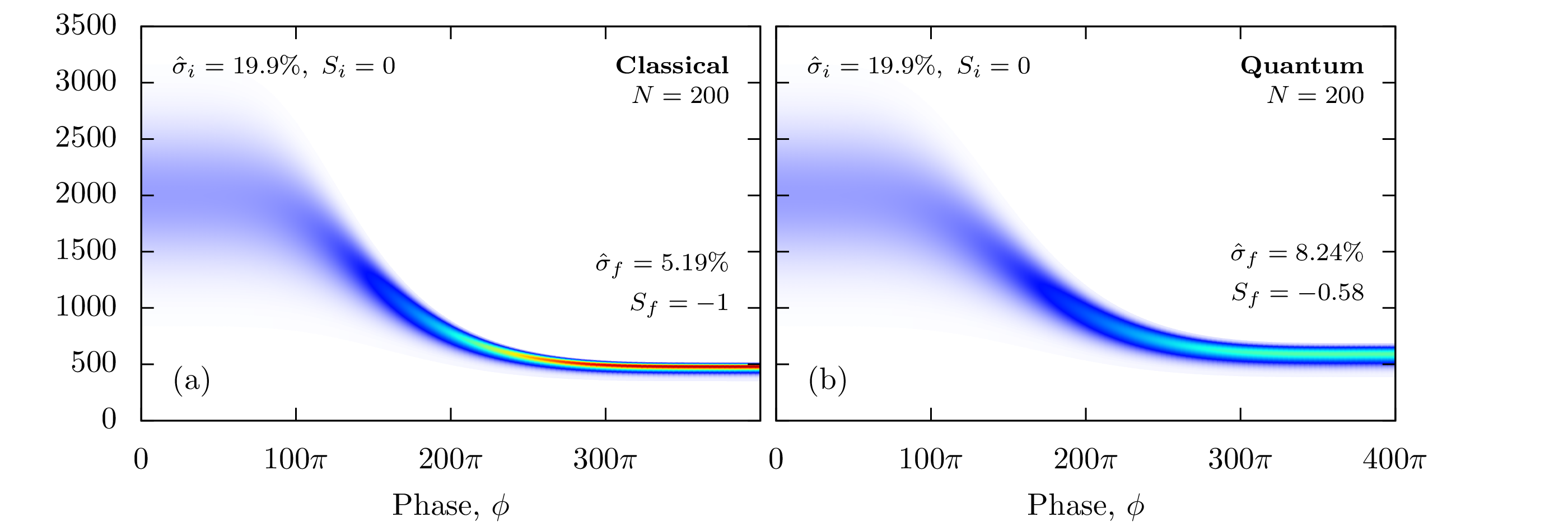}\\
  \vspace*{1em}
  \includegraphics[width=0.94\textwidth, clip=true, trim=0 70 0 0]
  {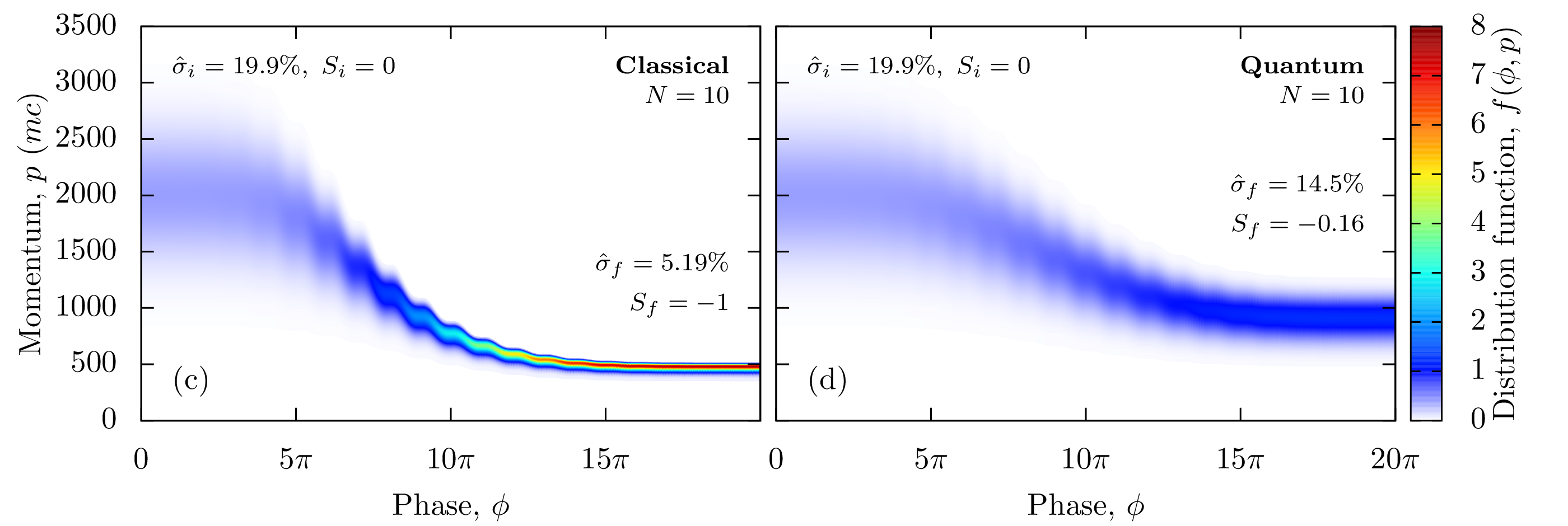}\\
  \vspace*{1em}
  \includegraphics[width=0.94\textwidth]{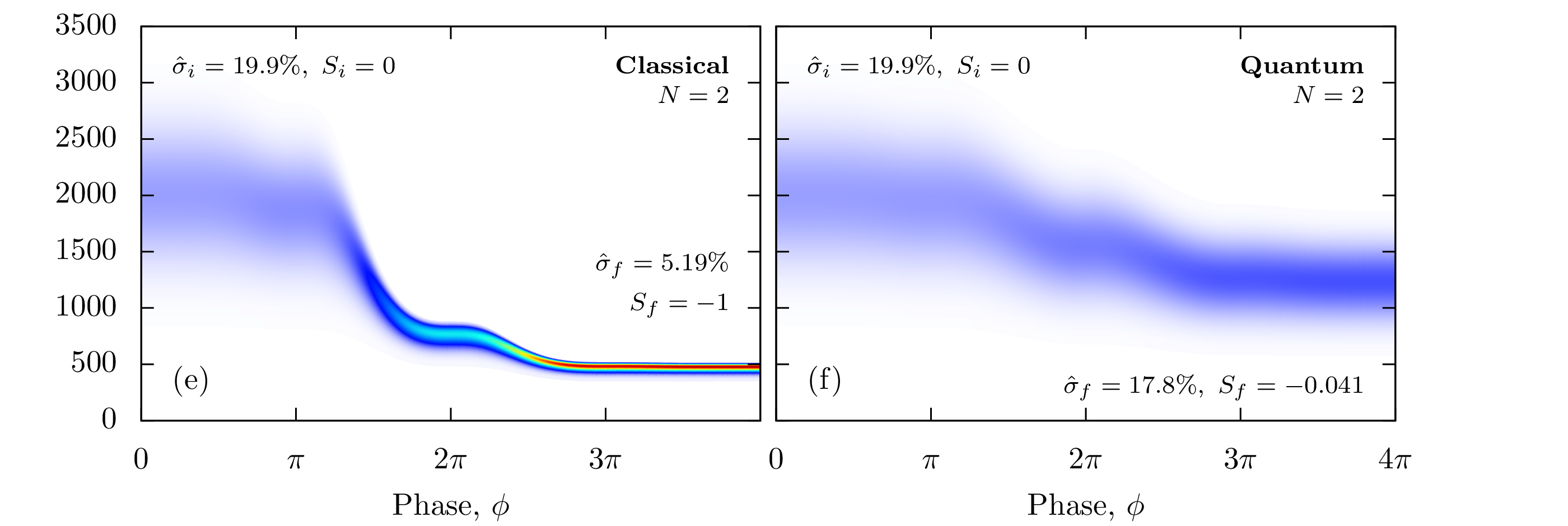}
 \end{center}
 \caption{(Color online) The phase space evolution of the distribution
 function $f(\phi,p)$. Classical predictions are shown in parts (a), (c)
 and (e), while the corresponding semi-classical results are presented in parts
 (b), (d) and (f). Values of the initial and final relative momentum
 spread and momentum skewness are displayed in each figure. The pulse length
 is reduced from $N=200$ cycles in (a) and (b), to $N=10$ cycles in (b) and
 (c), and finally to $N=2$ cycles in (e) and (f). In each case, we observe
 an increase in the final mean momentum and its spread predicted by the quantum
 model.}
 \label{fig:high_fluence}
\end{figure*}

To motivate this study, parameters have been chosen to be
comparable to those predicted for the forthcoming ELI facility. We have
chosen to consider $N a_0^2 = 4.624\times10^{4}$ which, for $N=10$ with a wavelength of
$\lambda = 800$ nm, represents a $27$~fs pulse duration \footnote{Note that the full-width half-maximum duration for this pulse shape is half of this value.} with
peak intensity $1\times10^{22}$~W/cm$^2$. We have investigated collisions
between pulses of length $N \in [2,200]$
cycles (together with their corresponding $a_0$) and a bunch of $\Np = \np$
particles, with an initial momentum spread of $20\%$ around $\bar{\gamma}
= \sqrt{1 + \bar{p}^2} = 2\times 10^3$. This corresponds to an average particle
energy of just over 1 GeV, which should be within the capabilities of the linear
accelerators at ELI.

Figure~\ref{fig:high_fluence} shows the variation of the particle
distribution function on the $(\phi,p)$ phase space. As can clearly be seen
in moving from the classical Landau--Lifshitz theory (left) to the quantum
model (right), there are noticeable differences in the mean $\bar{\v}$,
spread $\hat{\sigma}$, and skewness $S$ of the distribution.
We first note that the
difference in measuring the initial value $\hat{\sigma}_i = $ \SI\% $ < 20\%$
is due to
the finite number of particles used to represent the distribution.
Essentially, it comes down to the
`$\simeq$' in equation \eqref{eq:momentum_interval}, compared to the
definition given by equation
\eqref{eq:reconstruct_f}. As $\Np$ is increased, the approximation in
equation \eqref{eq:momentum_interval} improves, and the measured value
approaches the desired value. The value $\Np = \np$ was chosen as it allows
the error in the initial spread to be less than \initialerror\% (see
Supplemental Material). In practice, good
agreement can be found with lower $\Np$, with the caveat that properties
sensitive to the tails of the distribution (such as skewness) may be
strongly affected.

For the
classical theory, the final distribution only depends on the fluence
contained by the pulse, though this does not
prevent the system from taking different routes along the way. As we
decrease the number of cycles we observe in Fig.~\ref{fig:high_fluence} very different intermediate
behaviour, yet the measured properties of the final distribution support
this prediction: in each case, we measure the mean momentum $\bar{\v}_f =
\cp$ with a relative spread $\hat{\sigma}_f = \cs\%$. This represents a
significant contraction of the phase space, where the
average energy of the particle bunch has significantly decreased, as has its thermal
spread (beam cooling), and the distribution has become more sharply peaked.
In addition, we find the development of a negatively-skewed distribution
with $S_f = \cS$. In the classical model, the higher a particle's momentum the more it
radiates. This causes particles in the positive tail of the distribution
to be damped and slowed down more than those in the negative tail. We note
that the classical result has been verified by comparison with the analytic
solution of the Vlasov equation including the Landau--Lifshitz description of
radiation reaction \cite{Noble2014}.

The introduction of a semi-classical model in which the effect of radiation
reaction is reduced by the function $g(\chi)$ given by equation
\eqref{eq:QM_g} causes a reduction in the phase space contraction.
Figure~\ref{fig:high_fluence}(a) and (b) for $N=200$
clearly indicates this point, with the
final average momentum $\bar{\v}_f = \qap$ slightly higher than the classical case.
The final relative momentum spread is now \qas\%, showing that the final
distribution is no longer quite so sharply peaked. While remaining negative,
the skewness has also reduced in magnitude to $\qaS$, since it is
precisely the higher-energy particles which were previously most affected by
radiation reaction that now have this damping suppressed by larger $\chi$
(smaller $g(\chi)$).

These changes become even more pronounced as we move to higher intensities
(by reducing the number of cycles). For $N=10$, as shown in Fig.~\ref{fig:high_fluence}(c) and (d), we find that $\bar{\v}_f =
\qbp$ and $\hat{\sigma}_f = \qbs\%$ have both increased, with the skewness
also increasing to $S_f = \qbS$. This trend continues to $N = 2$
as shown in Fig.~\ref{fig:high_fluence}(e) and (f). In this case,
there is very little beam cooling for the quantum model, with the final
relative momentum spread taking the value $\hat{\sigma}_f = \qcs\%$
around $\bar{\v}_f = \qcp$. The profile also remains much
more Gaussian, with $S_f = \qcS$.

The reduction in 
phase space contraction
observed here is in
agreement with previous predictions \cite{Neitz2013}. This has allowed us to verify our
method for reconstructing the particle distribution function and enabled the
effects of the semi-classical model on a \emph{distribution} of particles to
be investigated. The distributions have been nicely reconstructed and do not
feature any artefacts or lack of resolution, unlike some other approaches
\cite{Thomas2012}.

\begin{figure}[!t]
 \centering
 \includegraphics[width=0.47\textwidth]{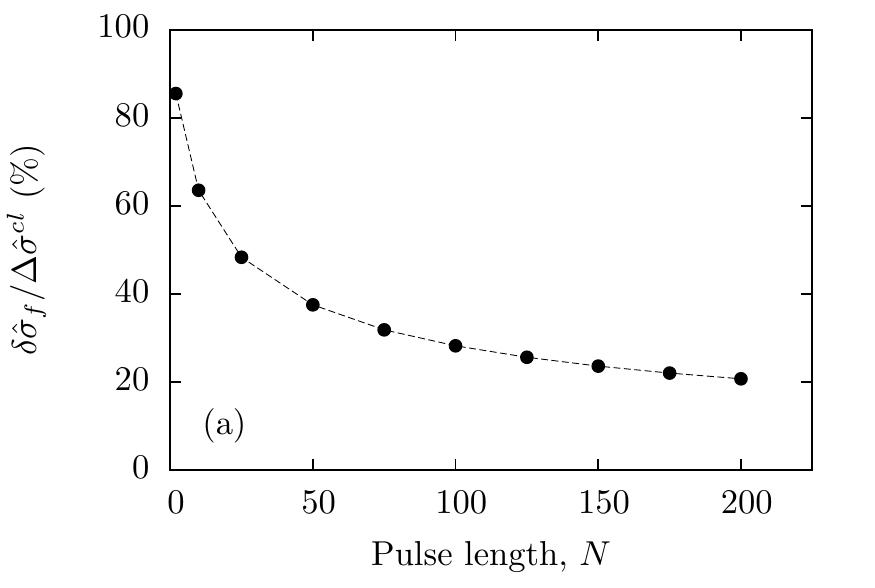}
 \includegraphics[width=0.47\textwidth]{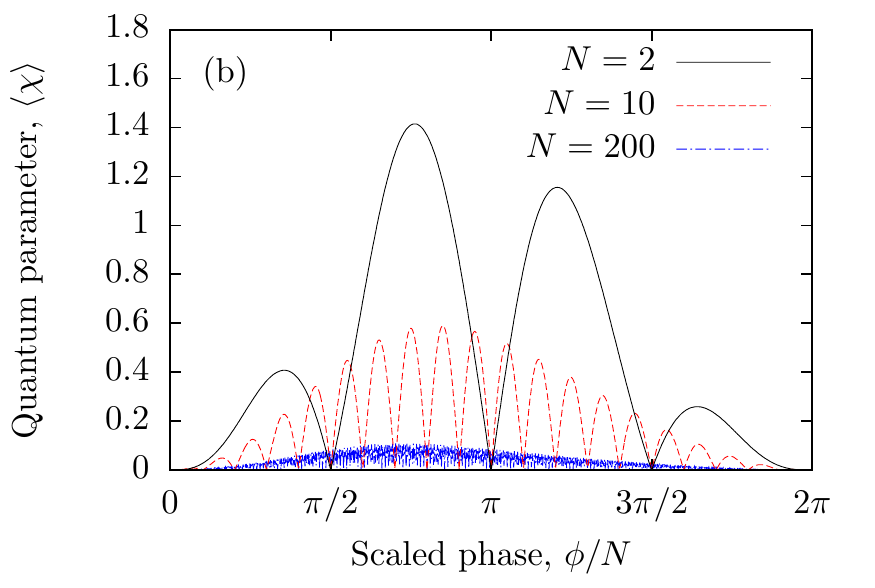}
 \caption{Part (a): variation of the final relative momentum spread difference
 $\delta\hat{\sigma}_f$ as a percentage of the total classical change,
 $\Delta\hat{\sigma}^\text{cl}$. The
 evolution of the average quantum parameter $\langle\chi\rangle$ is plotted
 in part (b) with phase scaled by $N$ in order to overlay the results. For $N \geq 10$, we find
 $\langle\chi\rangle < 1$ such that the semi-classical model should be
 valid.}
 \label{fig:high_fluence_validity}
\end{figure}

Figure~\ref{fig:high_fluence} also shows how the difference between the
classical and quantum results increases as we increase the intensity (or
decrease $N$). It is therefore interesting to consider the difference
$\delta\hat{\sigma}_f = \hat{\sigma}^\text{qm}_f - \hat{\sigma}^\text{cl}_f$
as a fraction of the total (constant) classical change in momentum spread,
$\Delta\hat{\sigma}^\text{cl} = \hat{\sigma}_i - \hat{\sigma}^\text{cl}_f$.
This can be found in Fig.~\ref{fig:high_fluence_validity}(a), where we see
that for $N=2$ the two predictions differ by around 85\%. As $N$ is
increased, this ratio is reduced as the average quantum parameter
$\langle\chi\rangle$
takes smaller values and radiation reaction is not so heavily suppressed. It
would be expected that the two models converge as $N\to\infty$.

In cases where the variation between the predictions of the two theories are
large, it is
important to be confident that the model remains valid. As a semi-classical
model, we expect it to remain valid into the weakly quantum regime, such
that particles experience instantaneous values up to $\chi \sim O(1)$.
Figure~\ref{fig:high_fluence_validity}(b) shows the evolution of $\langle\chi\rangle$
as the bunch moves through the laser pulse. The values plotted are for the
quantum model, since they remain smaller in the classical model. The cases
$N=10$ and 200 clearly remain below this threshold. The peak value $\langle\chi\rangle
\sim 1.4$ observed for $N=2$ is perhaps on the large side, but for such a
short pulse the plane wave approximation may be considered questionable
anyway.

In this Letter, we have introduced a novel method to efficiently and
accurately calculate the distribution function for an electron beam
interacting with an intense laser pulse. Using this method, we have compared
classical and quantum predictions of radiative cooling, finding that quantum
effects can significantly alter the beam properties, and unlike the
classical case can be sensitive to the shape of the laser pulse.

The results presented in this Letter are limited to the semi-classical case
$\chi \lesssim 1$. However, it should be noted that this restriction is due
to the use of a deterministic equation of motion, and not the method
of sampling and reconstructing the distribution. There should be no
obstruction to using this approach with a stochastic equation with photon
emission probabilities determined by strong field QED, as in
\cite{Green2014}, to explore strongly quantum regimes. This will be
addressed in future work.

This work is supported by the UK EPSRC (Grant EP/J018171/1); ELI Project;
and the 
Laserlab-Europe (Grant 284464) and EuCARD-2 (Grant 312453) projects.

\bibliographystyle{apsrev4-1}
\bibliography{refs}

\end{document}